\documentstyle[11pt]{article}
 
 \newcommand{\Msun}{M_\odot}
\title{\vspace{4cm}\Large\bf ON THE ORIGIN OF CIRCULAR
AND HEXAGONAL FORMATIONS IN GALAXIES}
\author{Yu.N.Efremov\\
 Sternberg Astronomical Institute, Moscow University, Moscow, 119899,
Russia,\\
}

\date{~}

\begin{document}

\maketitle

\begin{abstract}
\noindent
{\bf

The round and arc-shaped formations are known in some galaxies,
the Bubble complex (the Hodge object) in NGC 6946 being the most
remarkable. The rim of the complex has a form of a regular arc in
which a part of a hexagonal structure is embedded. A similar morphology
is recognized in  NGC 7421 galaxy, in that part of its rim which is
leading in the galaxy motion through the intergalactic gas, as suggested
by the bow-shock appearance of the HI halo of the galaxy. A hexagonal
shape is also found in NGC4676A galaxy. The HII radial velocities across
the Bubble complex  are compatible with its retrograde rotation and
drift, which are characteristic for solitary vortices known in nonlinear
hydrodynamics.  The drift motion may explain the location of the Bubble
complex at the tip of the largest elliptical HI hole in NGC 6946.
The hexagonal vortices in the atmospheres of Jupiter and Saturn are within
the gas streams what seems to be suggestive as well. It is conjectured that
the hexagonal rims of stellar systems might be relicts of flat segments
of  the shock wave produced by the ram pressure. The giant stellar arcs in
NGC 300 and M33 are associated with the energetic X-ray sources P42 and X-4
respectively (fig. 11), and these  might  be the relics of Hypernovae.}

\vspace{0.3cm}


\end{abstract}

\section{Introduction}

Most stellar systems  concentrate to the center of mass and have more or less
roundish  overall shape, without sharp rims (excepts the shell galaxies
and some other products of the galaxy interactions). It is natural,
for their shape is mostly determined by interplay of gravitation  and
rotation (internal motions). However there are stellar complexes
with sharp and very regular, arc-shaped rims, which  might have
a peculiar origin  (Efremov 2001, 2002). Some  galaxies are known
also to have sharp circular rims or the polygonal shape.  Here we  give
examples of such geometrically regular systems and discuss hypotheses
on their origin.

\section{The peculiar stellar complex in NGC 6946}

The Bubble complex (the Hodge object) in NGC 6946 was discovered
by  Hodge (1967) and independently by Larsen and Richtler (1999).
Its  origin is a subject of discussions.  The complex is really unusual
(Fig. 1), especially because its western rim is sharp and has the
shape of the regular arc with radius ~300 pc. The oldest stars are ~30 Myr
old and the rate of star formation there is 1-2 order of  magnitude higher
than that in the Local complex (Gould Belt). Another peculiarity of the
complex is the very massive young cluster near (yet not at) the center,
quite similar to ones known in the interacting galaxies (Larsen et al.,
2001; Larsen et al., 2002).

The complex rim is  sharp and circular only at the West, what was
the motivation to suggest the origin of the complex under the impact
of a cloud moved along the oblique trajectory from East to West
(Efremov, 2002). The highest light absorption is also observed along
this rim (Larsen et al., 2002). Other suggested scenarios were
the extreme case of the normal star formation (Elmegreen et al., 2000)
and the triggering by a Super-supernova  (Hodge, 1967).

The kinematics of HII gas inside and near the complex was studied
with the long slit at the 6m telescope (Efremov et al., 2002).
The radial velocity curves along all three  slit positions
demonstrate many perturbations, some of which may be interpreted
as the expanding shells or semishells. The most evident feature was
considered to be the semishell
with the velocity of expansion about 120 km/s and size of about
300 pc, its center is 7" East of the massive cluster.
Apart from these disturbances, there is the  wide positive bump
encompassing all the complex, with the average velocity about 20 km/s
larger than that of the surrounding field. It is worth noting that
this velocity is quite close to the velocity of the young massive cluster,
which was determined independently at 10-m and 6-m telescopes and is 150 km/s
(Larsen et al., 2001; Efremov et al., 2002).

We conjectured that the peculiarities of the velocity curves
which  are seen in Figs. 4, 5 and especially Fig. 6 in the paper by
Efremov et al. (2002) may be explained by the suggestion that
the gas motions in the complex are mostly connected with  vortices,
whereas the positive bump might  reflect the bulk drift motion of
the complex in the plane of the galaxy. The velocity curve
for PA -37$^o$ and the Western (undisturbed) parts of the curves
for PA 83$^o$ and PA 29$^o$ demonstrate  the linear gradients, which  are
different for different position angles in a way compatible
with the solid body rotation in the plan of NGC 6946,
to the direction opposite to the rotation of the galaxy.

The retrograde rotation and drift are expected for the Rossby soliton in
the galaxy disk  (Korchagin and Petiashvili, 1985; Korchagin and
Ryabtsev, 1991). Considering the positive bump of HII velocity
has the same value as the main cluster velocity and both deviate from the
rotation of the galaxy, the prograde  drift of the complex seems to be a
possible interpretation. The amplitude of the wide bump being about 20 km/s in
radial velocity and and the inclination of the galaxy plane to the sky
being  32$^o$, this translates to the velocity in the plane of
the galaxy about 40 km/s.

It is possible that the drift of the complex  in relation to the
stuff of the galaxy disk may explain the strange absence of
the HI hole around it. Moreover, overlay of the Kamphuis (1993)
map of HI hole in NGC 6946 and the galaxy image from the DSS
(Fig. 2) demonstrates the position of the complex  at the Southern
tip of the supergiant  elliptical ($6.5 \times 2.9$ kpc) HI hole
(\#13, surely) outside the latter.
The geometry of the situation, the direction and velocity of the suggested
complex drift imply that 30 Myr ago (the age of the first burst of star
formation in the complex)  the complex was located well inside the hole
\#13. One may suggest that formation of the hole \#13 and the
complex were triggered by the same HVC impact. This hole is associated with
the high velocity gas region \#12 (Kamphuis, 1993, Table 2) which has
energy $8 \times 10^{53}$ erg and mass $3.1 \times 10^7 \Msun$,
whereas the missing mass in the hole \#13 is $22.3 \times 10^7 \Msun$
(Kamphuis, 1993, Table 1). The oval shape of the hole \#13
may reflect the way of the Bubble complex  through the gaseous
disk  and the present day location of the complex near the hole tip
(yet outside it) may  indicate its recent arrival to this point
and the time needed to form the expanding HI hole around.

Our hypothesis implies that under some conditions a cloud impact may
trigger the  gaseous vortices in the disk of the galaxy.
There is no large scale perturbation of the NGC 6946 galaxy rotation curve
needed in the theory to trigger the solitary vortex (Rossby soliton).
We suggest a special origin of a vortex, not connected
with the bulk dynamics of the galaxy.  The loop of the velocity field,
seen in the region of the Hodge object  (Bonnarel et al. 1988)
may be result of the initial local perturbation.
The  vortices indeed formed in the tail of the impacting cloud
in the Santillan et al. (1999) models of the HVC impact to the magnetized
galactic gas disk. May such gas motions trigger then the star formation?
At any rate, the  star formation might be  confined within the long
standing gas vortex.

The radial velocity curve for PA 83$^o$ (Figs. 4 and 6 in Efremov
et al. 2002) contains the deep dip which was explained in that
paper as the expanding semishell. The wide positive bump is
centered on the massive cluster, yet the dip is in the center of
the additional more narrow positive bump to the East of the
complex center. This configuration must be explained. One may
suggest that  here we have not the semishell expanding to us  yet
two adjacent vortices with very fast and opposite rotation,  the
situation expected in cosmic hydrodynamics and known as von
Karman vortice street (Chernin, 1996).

The expanding semishell might be connected with the position of
the pressure source outside the plane of symmetry  of the
galactic plan (what is compatible with the Hypernova hypothesis
for the  semishell origin, Efremov et al. 2002).  However,  the
plot of $V_r$ for such a structure  should be a sinusoid, whereas
we rather see two straight lines, especially  certain for the
Eastern  side of the $V_r$ dip. These two lines suggest existence
of two  vortices,  rotating in opposite directions. 
They  might be parts of "von Karman vortice
street" in the tail of the impacting cloud, such as ones seen in
model by Santilan et al. (1999).  Such vortices are not the full
rings and, apart from this,  the rotation in the the galactic
plane easy  explains why no line splitting is seen over the dip.

\section{Ram pressure and star formation}

The best way to understanding a peculiar property  is to find
other examples of  systems which share the property.
So we have been looking for circular rims
everywhere. Indeed,  130-degree long arc  of the Western rim
of the Bubble complex has analogies in the  the Quadrant and Sextant
arcs in the LMC, yet also in rims of a few galaxies.

The galaxies with  the circular rims are known, indeed.  It is DDO 165
in the M81 group, the Southern rim of which is sharp and arc-shaped
(Efremov,  2001).  In the same M81 group,  the Ho II galaxy has the
outer HI isolines of the characteristic comet-like shape
(Bureau and Carignan, 2002),
and  SE part of HI halo of this galaxy is bordered by the perfect arc
of a circle. These authors suggest  the action of the ram pressure
as the  reason for this shape.  Even earlier the same suggestion was
proposed by Ryder (1997) to explain the similar shape of HI halo of
NGC 7421 galaxy and it is surely true for NGC 2276 galaxy
as well (Fig. 3).

The side of the stellar disk of NGC 7421 turned to direction of
the movement is evidently enhanced  and sharp-bordered. This is even
more evident for NGC 2276 galaxy, where the HII isolines
(Gruendl et al. 1993) and even the stellar disks are shaped in
accordance  with  the Mach-cone like appearance (Fig. 3)
of the galaxy at 1.49 GHz (Hummel and Beck, 1995).

These are the clear evidences for the star formation
triggered by  the ram pressure  and still preserved
the characteristic bow-shock shape.
We may therefore suggest that the southern rim  of DDO 165 (fig. 4)
is also relict of the star formation triggered
by the ram pressure in the bow-shock shaped edge of the galaxy gas
(Efremov, 2002).

The conclusion follows that the 130-degree long western arc
of the Bubble complex in NGC 6946 may also be relict of the star
formation in the bow shock at the interface of the impacting  cloud
and NGC 6946 gas disk.   The same explanation may be true also for
the LMC arcs in the LMC4 region, their occurrence in the same (NE)
region  of the galaxy being explained  by observation that
the NEE  edge of the LMC is leading  in the LMC motion and
is the first to meet the clouds of the Galactic halo (Efremov, 2002).

We noted that this may be confirmed by the bow-shock appearance
and orientation of the X=ray emission
around SNR N63  (Chu, 1999) localized at NE of the LMC4 supershell,
near the NE edge of the LMC gaseous disk (Fig. 5 and 11)  - it may be shaped
by the ram pressure owing to the LMC  motion through the halo of
the Galaxy.    The line drawn through the X-ray maximum  under
the position angle of the LMC proper motion 79$^o$
(Marel et al., 2002)  is close to the axis of the symmetry of X-ray nebula
(Fig. 5). The sharp NE edge of the distribution of HI in the LMC
is in accordance with this.

\section{Hexagonal rims connected with ram pressure?}

The close examination of the Bubble complex rim under different image
contrasts reveals that  at least half of its rim is confined
by segments of  hexagon  inscribable into a circle.  This is seen
for the "leading" (Western) part of the complex in the HST images, whereas
in NOT images all the complex, after application of the non-monotonic
characteristic curve, display  the pentagon shape (Fig. 1).

There are also the examples of galaxies with hexagonal or pentagonal rims.
One may note that the leading rim of NGC 7421  is  also
outlined by the segments of hexagon inscribable into the arc of the circle.
Something like this is seen also in NGC 2276 image yet not so
definitely  (Fig 3). The  similarity of the leading edge rims
of these two galaxies and the western rim of the Bubble complex
is evident. These galactic rims  are certainly shaped (in the end)
by the ram pressure  and that of the Bubble complex is plausibly too.
Anyway, a number of other galaxies (especially in the Virgo cluster)
are known to  have signatures of the ram pressure action yet have
no the hexagonal rims. This may be explained by observation that
the galaxy must be seen pole on, in order the  polygonal shape
of its disk would  be noted, as it is the case for NGC 7421 and NGC 2276.
Otherwise, the hexagonal shape may be a transient property of the
ram pressure shaped gaseous cloud, which may not be necessarily
preserved in the stellar distributions.

Other astronomical observations may be the clue to the issue.
There is known at the Saturn North pole the hexagonal cloud
feature discovered in Voyager close-encounter images (Allison et
al., 1990). These authors interpreted the hexagonal cloud as a
stationary Rossby wave. They note also that the wave is embedded
within a sharply peaked eastward jet (of 100 meters per second)
and appears to be perturbed by at least one anticyclonic oval
vortex immediately to the south.

In result of browsing of  available images of the Jupiter
Red spot we have noted  that sometimes the vortex
bordered by segments of hexagon is developing inside the Spot, while
the leading rim of the Red spot has acquired the shape of two straight
lines, as it is seen in the image obtained by Voyager in 1979 (Fig. 6).
The turbulent vortices behind the Spot are well seen too
(see http://antwrp.gsfc.nasa.gov/apod/ap020205.html).

As the hexagonal cloud at Saturn, the Red spot
is known  to locate in the stream of the fast moving gas of the Jupiter
atmosphere, yet it is the persistent vortex not sharing the movement
of encompassing gas.  The analogy with the ram pressure action to
the galaxy moving through intergalactic gas is evident.

The experiments with rotating shallow water demonstrated
the arising of the structure interpreted as the Rossby solitary
vortex, which emerges because of the self-organization of dissipation
system far from the equilibrium state  (Wang et al. 2001). These authors
concluded the experiment is a model of the Jupiter Red Spot.
In fact, the similar experiments were carried out by many authors
(see the references in Korchagin and Ryabtsev, 1991), and the suggestion
that Jovian Red Spot is a stationary nonlinear soliton-like vortex
was advanced long ago by Maxworthy and Redekopp (1976).
It is worth noting that, as was demonstrated by
Korchagin and Ryabtsev (1991) the hydrodynamical theory of solitary
vortices is applicable to a stellar galactic disk as well.

The hexagonal and asymmetrical galaxies are seen also in the
Hubble Deep Fields. Some samples extracted from the ingenious
site of S.Gwyn   

(http://astrowww.phys.uvic.ca/grads/gwyn/pz/hdfs/spindex.html)
are given in Fig. 7. These galaxies are mostly in small groups
and the ram pressure origin of the hexagonal shape is plausible.
It is worth noting that in some galaxies (293N and 222S) there
are twin blue spots at the presumably rear (in motion) edge.
These spots might be the result of the star formation in the
vortex couple known to locate behind the fast moving body (see
also Fig. 1).

It is possible that the hexagonal shape
might be the transient property of the persistent vortices,  which are
objects of the ram pressure.  It is a result of self-organization,
reminding by the resulting shape (yet not by physics of origin) the cells
of Benard. If the star formation occurred in the shock wave while its
front was flat, the facing to the stream  part of the stellar
system rim would preserve the hexagonal shape.

Anyway, note the Benard-cells like appearance of the unique
Honeycombe feature in the LMC, located at ~100 light year from
the SN1987A. The structure consists of two dozen adjacent gaseous
cells, most of similar size ~ 3 pc (Noever, 1994; Meaburn et al.
1995) and the high velocity gas motions observed along the cell
walls (in direction perpendicular to the cell planes) remind the
convective motions along the walls of the Benard cells. Redman et
al. (1999) suggest  that the cells were formed owing to the
Rayleigh-Taylor instability in the shell of the old SNR, whereas
Noever (1994) noted that regularities of the polygons of the
Honeycomb  follow some laws of the statistical crystallography.

Chernin (1996) noted that it must be explained why the largest known
eddies in the Universe are galaxies,  and one may wonder if the whole
galaxy might be considered  as the Rossby soliton.  A far going idea of
such a kind was published, indeed (Dubrovsky, 2000).  In accordance with
this suggestion, the rotational velocity of a galaxy  is determined by
the vortex instability rate and not the central mass, so the hidden mass
may not be necessary.

\section{What are the dark rings?}

The dark rings  surrounded by arcs of clusters are
known in  M83 (Efremov, 2001) and  NGC 4449 (Bothun, 1986), of
about 500 pc diameters, like the Hodge complex  (Fig. 8). They are seen
in some other galaxies too  (NGC 6946,   NGC 2207).
At the first glance, these rings  might be considered
the examples of the star formation triggered by the gravitational
instability in the swept up gas shell, more so the distances between clusters
in M83 Western ring  being remarkably equal.  However, a  swept up
expanding shell should leave the inner space of the ring empty from gas and
dust, whereas there is dust; the central clusters are  not seen
in M83 rings (the bright object inside the Southern ring should be
the foreground star), yet are in NGC 2207 and NGC 6946 (Fig.8).
The visible darkness of the NGC 4449 complex inside the ring is confirmed
by the photometric data (Hitchock, Hodge 1968).
If so, the gas density  inside such rings is higher than in surroundings
and the circular shape may suggest these rings are gaseous (inspiralling)
vortices with the enhanced column gas density at least along their rims,
where the star formation was effective.

The multicolor observations
could find the parameters of the light absorption  inside these rings.
It would not be surprising to find these to be unusual, like it is the case
for the dark cone centered on the star cluster in the highly peculiar
stellar complex in NGC 2207 (Elmegreen et al. 2000  - see Fig. 8).
The very high  resolution HI and CO observations  could also solve
the problem. To find the real absence of
even old stars  inside the  dark rings would be tremendous yet surely
improbable possibility.  The features of this kind found in the cores
of a few galaxies are much smaller (Lauer et al., 2002)

Considering that inside the dark rings in NGC 6946 and NGC 2207 the
central objects (clusters?) are seen (and there is a guess for this
in NGC 4449, Fig. 8), these rings may still be examples of the
swept up shells, after some  interactions of the shock waves.
The smaller scallops at the inner side of both rings in M83 (the only
images with good resolution) may also suggest this.    Anyway, the
known  tested examples of the star formation in  the swept up shells,
such as in IC 2574, are quite non-similar to these rings, as well
to other formations discussed here (see Efremov, 2002).

\section{Other hexagonal  galaxies}

The hexagonal shape is well known for the rings surrounding
some galaxies and also for bulges of some S0 galaxies.
The list of such galaxies and discussion of their nature was recently
given by Chernin et al. (2002).
They noted that the galactic rings are intimately related to spiral
structure, as   Buta and Combes (1996) demonstrated, and the origin of
the hexagonal structure may be the same as of the polygonal arms known
in many grand design galaxies (Chernin 1999, Chernin {\it et al.} 2000).

As Chernin et al. (2002) noted, there is  the major similarity
between hexagon structures  and polygonal shape
of the spiral arms: in both structures  the angle between adjacent
straight lines is about $2\pi/3$ and the lengths of the straight segments
are equal to the radius of encompassing circle  or to the local radius
of curvature of the logarithmic spiral. Chernin (1999) suggested
that the local flattening
of the spiral front on the space scale of local radius of curvature may
be due to stability of a flat shock wave against any weak perturbations
that disturb its front surfaces. On this basis, Chernin et al. (2002)
concluded that the hexagons are also made of flat segments of shock
fronts.

This suggestion may explain the   hexagonal-shaped rim of the Bubble complex
and of the galaxies which, like NGC 7421 and NGC 2476, are object of the ram
pressure. The occurrence of the straight segments of rims of these galaxies
at the shocked  side only is especially indicative.

However,  there are  rare galaxies which are hardly objects
of the ram pressure yet whose disks have hexagonal shape. The long known
example is  NGC 6776 (Sansom {\it et al.} 1988) which is considered  to be
E galaxy but is certainly S0, being rotating.  Another is NGC 1637,
the three-armed well isolated spiral (Fig. 9).  There is no
other galaxies within  1$^o$ from it (Block et al. {\it et al.} 1988).

Another problem is the peanut shape of bulges of some S0 galaxies,
the best samples being NGC 7020  (Buta, 1990) and IC 4767 (Whitmore and
Bell, 1988). These galaxies look like the inclined hexagons.The proposed
explanations were  the interactions with other galaxies
or specific shapes of stellar orbits. The  key issue here is
the orientation; the visible prolate hexagons may be intrinsically
prolate thick bulges seen edge-on or  they may be the hexagonal disks
seen under angle. The former possibility seems to be easier explainable
by the  orbit orientations  for the NGC 7020 case  (Buta 1990).
Also, Merrifield and Kuijen (1999) argued that the boxy and peanut-shaped
bulges of some galaxies are galactic bars viewed from the side.

We found an example of the hexagon where the edge-on orientation is sure.
It is  NGC 4676 ("the Mices") interacting galaxies case. The image
obtained recently by the  HST with the Advanced Survey Camera
(see http://oposite.stsci.edu/pubinfo/pr/2002/11/)
after editing  reveals the hexagonal shape
of the Northern (NGC 4676A) galaxy, whereas  the existing kinematical
data indicate it is  seen  edge-on (Yun and Hibbard, 2001).
This inclination follows also from the thin and straight appearance of
the long tail of the NGC 4676A. The tidal tail should be intrinsically
wide and planar, so in the Mices system it is seen edge-on.
The overlay of CO isolines from Yun and Hibbard (2001)
and the HST ACS image (Fig.10) demonstrates that  not all dust
clouds seen in NGC 4676A are connected with CO emission in the galaxy
and vice versa.  This might be partially explained by the localization
of some dust clouds in front of the galaxy body.

These interacting galaxies are in the Coma cluster and may be objects
of the ram pressure. We suggest that the curved cone-like plume at the North
of NGC 4676B is just the manifestation of action of the ram pressure onto
the leading edge of the galaxy. Overall structure of the system
may  display the orbital motion of the  Mices.
The blue color of the turned cone tip  of NGC 4676B
is surely result of the star formation, triggered by the ram pressure.
If so, one may still wonder if the ram pressure was participated
in shaping the NGC 4676A galaxy as well.

\section{Stellar arcs and Hypernovae}

The best known arc-shaped complexes are the LMC4 group
of arcs 200 - 300 pc in radii, noted first by Hodge (1967).
The concentration of arcs near to each other (Fig. 11) was explained
by their origin from  the objects, ejected from the massive 2 Gyr old
cluster NGC 1978 in the same area.  It was suggested
that  the progenitors of GRB/Hypernovae  are binaries of the compact
objects  formed owing to dynamical interaction of stars and/or their
remnants within the dense cluster and then ejected from it.  We argued
that this is the main channel of formation of binaries with compact
components (Efremov, 2001b).

The same idea is now  accepted to explain the concentration
of X-ray binaries  to  the globular clusters (White et al. 2002).
According Kundu (2002), 40 percent of the brightest LMXBs (which are suggested
to be the black hole accretors) in the elliptical galaxy  NGC 4472
are associated with the globular clusters.  We have noted the concentration
of the X-ray binaries in the region of the LMC4  (Efremov 2001b and
references therein) and may now conclude that idea on the origin of
progenitors of the LMC4 arcs in the NGC 1978 cluster is still alive.

The origin of the LMC system of arcs due to the ram pressure onto the
surface of the impacting cloud was also considered (Efremov, 2002).
The physically similar situation arises after interaction of a dense
and cold enough cloud  with the blast wave from  the  powerful
external explosion.  The large increase in pressure leads to
the compression of the cloud, most rapid at the face, exposed
to the blast wave, and the bow shock may appear along this side
(McKee and Cowie, 1975). The observational data discussed above
demonstrate that  the triggered star formation  may result, the bow shock
appearance being preserved in the distribution of the young stars.

The 400 pc in size arc of the bright stars, noted in Efremov (2001), and
now known as the complex AS102, in the spiral arm of the NGC 300  galaxy
may  be result of such an event (Fig. 11).
The brightest in the galaxy  X-ray source,  P42 = H13, which is
classified as X-ray accreting binary system  including a black hole
(Read and Pietsch, 2001), is near the complex at its convex side and
exactly on the axis of symmetry of the stellar arc (Efremov, 2002).
Considering the age of the complex (~5 Myr, Kim et al. 2002) and its
distance to P42, the energy needed to compress the paternal cloud
to AS102 stellar complex should be that of a Hypernova.

The last example is the arc of the OB-association HS137 in M33, noted
in Efremov (2001).  It is also known as the HII region IC133 (Fig. 11).
This arc encompasses the HI hole \#31, coordinates of which
are the same as of X-ray source M33 X-4 (Schulman and Bregman, 1995).
This source was investigated recently by  Okada et al. (2001)
who found it to be the young SNR with energy
considerably higher than those of SNRs in the Milky
Way.  This X-ray source is the only in M33 which is suggested to be
physically associated with a HI hole (Schulman and Bregman, 1995).

The arc-like appearance of the AS 102 complex formed by the outer
pressure to the paternal cloud is natural.  However, for IC133 complex
we evidently deal with the classic model of the star formation in
the swept up shell, and the stellar arc was formed just at the side
of the shell where the observed HI density is higher.  It was just
assumption by Westerlund and Mathewson (1966) to explain the "arc
of blue stars" in the LMC4 region now known as the Quadrant arc.
However, this suggestion does not work just to explain the LMC4 arcs,
for the new data demonstrated the Quadrant  is deep inside the LMC4
HI hole. The rather similar complex in M83  was suggested to arise
in the result of the HVC impact (Comeron, 2001), and this may be
the case for the LMC arc complex as well (Efremov, 2002).

\section{Concluding remarks}

Putting the things together, we hypothesize that  the hexagonal appearance
might be the transient  property of  persistent vortices (of wide
size range),  which are suffered from the ram pressure.
The straight segments of the rims might be a result of the
the local flattening  of the shock  front on the space scale of local
radius of curvature what  might, in turn, be due to stability of a flat
shock wave against any weak perturbations that disturb its front surfaces
(Chernin, 1999). If the star formation occurred in the shock wave
while its front was flat, the respective part of the rim of the stellar
system would preserve the hexagonal shape. The flatenning of the shock
wave fronts seen in galactic spiral arms (Chernin 1999) and now in other
formations worth to be investigated.  May be the Chandra data for clusters
of galaxies will find something of the kind.

However, the origin of most  round or arc-shaped complexes is
still a problem. Considering
that both stellar arcs in NGC 300 and  M33 are the only such
features in these galaxies and associated with the  most unusual
in the respective galaxy X-ray sources, both well isolated,  the
chance coincidence seems to be improbable. We believe that these
X-ray objects, P42 and X-4 (especially P42), will be proved to be 
the remnants of Hypernovae. The estimated young age of the M33 X-4 
SNR (Okada et al.,  2001) is surely incompatible with our suggestion, 
yet the source should be studied in more details. Both these arcs 
are rather  irregular, unlike  other arcs discussed here.

The similar origin is possible also for the LMC4 arcs, yet there are
no X-ray sources located at the suitable positions in respect to the arcs.
The X-ray binaries  concentrate  mostly to North of the LMC4 supershell,
near the NGC 1978 cluster. The origin of the LMC arcs in the result
of the ram pressure to the surface of impacting clouds seems to be more
probable, considering also  the similar opening angles of these
one-sided arcs. The main problem is that there are at least 2 and probably 5
arcs close to each other; it is a difficulty for any hypothesis, as well
as the different orientations and ages.  The Hodge object is enigmatic
also, with a couple of concentric semiarcs, suggesting action of a central
pressure  yet having nor an evident source, neither an age - space
pattern (see Fig. 1 here and  Fig. 9 in Larsen et al., 2002). The vortex
motions might confine the young stars within the round complex; however,
the sharp arc of the Western rim suggests the one-sided ram pressure action.
Note that the drift of the complex might explain the absence of the HI
hole around it, what is very strange considering the complex contains
the cluster most suitable to form a supershell. May be the understanding
of these features will come from an unexpected side,
such as the dark matter  presence, or something else like this...

One may wonder how frequent are the round peculiar formations in galaxies.
We were able to find  a dozen only  (Efremov 2001).
The Hodge object in NGC 6946 was the only result of the special
searches for the features similar to the giant arcs in the LMC4 region
(P.Hodge, private communication). The appearance of this complex under
different resolutions and contrasts suggests that many similar
features may not be remarked. Under the low resolution, and in
the distant galaxies they are practically  star-like
(the best examples are the round complexes in M51 and especially in
NGC 1232, see Fig. 12),
whereas under the low contrast (such as in the Sandage-Bedke atlas)
the encompassing circular rim is unnoted.  This is also the case for
the stellar arcs even so large as in the LMC4 region. The rather similar
in size complex in M83 (Comeron 2001) includes also two giant arcs
(Fig. 12), which are  seen only in the best resolution images (Efremov, 2001).

At any rate, these peculiar entities are curious and elegant, and
their interpretation may have the far reaching implications.
Some were suggested in this  paper.

\section{Acknowledgments}

I am very grateful to A.Chernin for the comments on the paper draft
and many useful discussions, and to P.Korchagin for comments
on the theoretical motions of a solitary vortex in a galaxy.
The  using of the NASA Astrophysics Data Systems, the LANL electronic
preprint service, the Digital Sky Survey, the US Naval Observatory
Flagstaff Station Integrated Image and Catalogue Archive Service,
and  also the HST and ESO Press Releases was necessary
to  carry out this work and is thankfully noted. Thanks are due also
to S.Larsen and I.Karachentsev for the images obtained  with the NOT
and the BTA. The  support from grants RFBR 00-02-17804 and  00-15-96627
is appreciated.

\vspace{1cm}

{\bf REFERENCES}
\vspace{0.3cm}

Allison M., Godfrey D. A., Beebe, R. F. (1990) Science, 247, 1061,

Block D.L., Puerari I., Frogel J.A. et al. (1999) Astron, Sp. Sci., 269, 5.

Bonnarel F., Boulesteix J., Georgelin Y.P. et al. (1988), A\&A, 189, 59.

Bothun G.D. (1986) AJ, 91, 507.

Bureau M., Carignan C. (2002) AJ, 123, 131.

Buta R., 1990, ApJ, 356, 87

Buta R., Combes F., 1996, Fund. Cosm. Phys., 17, 95

Chernin A.D. (1996), Vistas in Astr. 40, 257.

Chernin A.D., 1999, MNRAS 308, 321

Chernin A.D., 2000, MNRAS 318, L7.

Chernin A.D., Zasov A.V., Archipova V.P., Kravtsova A. S., (2000)
in: Galaxy disks and disk galaxies", eds. J.G.Funes and E.M.Corsini,
ASP Conf. Ser., Vol. 230, p. 147.

Comeron F. (2001), A\&A, 365, 417.

Chu, Y.-H. (1999), Ap. Sp Sci., 269-270, 441.

Dubrovskiy V. A. (2000) in:  New Cosmological Data and the Values
of the Fundamental Parameters, p. 52, IAU Symposium no. 201, Manchester.

Efremov Yu. N. (2001) Astron. Rep., 45, 769.

Efremov Yu. N. (2001b) astro-ph/0102161.

Efremov Yu. N. (2002)  Astron. Rep., 46, \#10 (in press).

Efremov Yu. N., Pustilnik S.A., Kniazev A.Y. et al., (2002) A\&A, accepted =

astro-ph/0205368.

Elmegreen B.G., Efremov Yu.N,, Larsen S.S. (2000) ApJ, 535, 748.

Elmegreen B.G., Kaufman M., Struck C. et al. (2000) AJ, 120, 630.

Gruendl R.A., Vogel S.N., Davis D.S., Mulchev J.S. (1993) ApJ, 413, L81

Hitchock J.C., Hodge P.W. (1968) ApJ, 152, 1067.

Hodge P.W. (1967) Publ. ASP, 79, 297.

Hummel E., Beck R. (1995) A\&A, 303, 691

Kamphuis J.J. (1993) Ph. D. Thesis, Univ. Groningen.

Kim S.C., Sung H., Lee M.G. (2002) astro-ph/0203032.

Korchagin V.I., Petviashvili V.I. (1985) Sov. Astr. Lett. 11, 121.

Korchagin V.I., Ryabtsev A.D. (1991) AA, 246, 368.

Kundu A. (2002) astro-ph/0206221.

Larsen S.S., Richtler T. (1999)  A\&A, 345, 59.

Larsen S.S., Brodie J.P., Elmegreen B.G. et al. (2001) ApJ, 556,
801.

Larsen S.S., Efremov Yu.N., Elmegreen B.G et al. (2002) ApJ, 567,
896.

Lauer T.R., Gebhardt K., Richstone D. et al. (2002) astro-ph/0206122

McKee C.F. and Cowie L.L. (1975) ApJ, 195, 715.

Maxworthy T., Redekopp L.G. (1976) Icarus, 29, 261

Marel van den B.P., Alves D.R., Hardy E., Suntzev N.P. (2002),
astro-ph/0205161

Meaburn J., Wang L., Bryce M. (1995) A\&A, 293, 532.

Merrifield M.R., Kuiken K. (1999) A\&A, 345, L47.

Noever D.A. (1994) A\&SS, 220, 65.

Okada Y., Takahashi H., Makishima K. (2001) PASJ, 53, 663.

Read A.M., Pietsch W., (2001) Astron. Astrophys., 373, 473.

Redman M.P., Al-Mostafa Z.A., Meaburn J. et al. (1999) A\&A, 345, 943.

Ryder S.D., Purcell G., Davis D., Andersen V., (1997) Publ. AS
Austr, 14, 81.

Sansom  A.E., Reid I.N., Boisson C. (1988) MN RAS, 234, 247.

Santillan A., Franco J., Martos M., Kim J., (1999) ApJ 515, 657.

Shulman E., Bregman J.N. (1995) ApJ, 441, 568.

Wang Z. P., Wang L. Y. Liu, S. S. (2001) Acta Astr. Sinica, 42, 397.

Westerlund B.E., Mathewson D.S. (1966) MNRAS, 131, 371.

White R.E., Sarazin C.L., Kulkarni S.R. (2002), ApJ, 571, L23.

Whitmore B.C., Bell M. (1988) ApJ, 324, 741.

Yun M.S., Hibbard J.E. (2001) ApJ, 550, 104.

\newpage

\section*{Figure captions}

Fig 1. The images of the Hodge object in NGC 6946.
At the top - the segment of the HST WF camera  image.
Note the symmetric arcs of clusters "behind" (at the left)
of the complex - the stellar relicts of von Karman vortex street?
At the bottom - the edited NOT image, with the encompassing circle added.
Note slightly different orientation; Nord is up in the NOT image.

Fig. 2. The map of HI holes in NGC 6946 (Kamphuis 1993) overlayed by
the galaxy image map from  USNO Image and Catalogue Archive Service.
The Bubble complex is near \#13, looking quite similar to the bright
foreground star.  North is up.

Fig. 3. The ram pressure shaped NGC 7421 (top) and NGC 2276 (bottom)
galaxies. At the left  HI (NGC 7421) and 1.79 GHz  (NGC 2276) isolines,
at the right - the DSS images.  North is up. See the text.

Fig. 4. The DDO 165 galaxy images from the plate of 6-m telescope,
courtesy of I.Karachentsev.

Fig. 5. The SNR N63 in the LMC  (Chu, 1999). The isolines of X-ray emission
are shown with the direction of the LMC proper motion added.  North is up.

Fig. 6. The Jovian Red Spot in 1979. See the text.

Fig. 7. The samples of galaxies from the Hubble Deep Fields,
probably shaped by the ram pressure. The top row - 120S and 293N,
the middle row - 222S and 41N, the bottom row - 255N and 96N. See the text.

Fig. 8.  The left row - the Western (top) and Southern (Efremov, 2001)
dark rings in M83 (the VLT ESO images); the NGC 4449 ring (bottom).
The right row - the dark rings in NGC 6946 (top, NOT image); the dark ring
in NGC 2207;  the peculiar complex, which includes the elliptical
dark ring, in NGC 2207 (bottom). The last two HST images are from
Elmegreen et al. (2000).

Fig. 9. The hexagonal NGC 1637 galaxy (the edited DSS  image).

Fig. 10. The Mices (NGC 4676AB) galaxies. It is the overlay of the HST ASC
image and CO isolines from Yun and Hibbard (2001).  North is up.

Fig. 11. At the top - the system of giant stellar arcs in the LMC4 region
(the edited image based on the Boyden Observatory photograph obtained by
H.Shapley; courtesy of P.Hodge and K.Olsen).
At the bottom left-  the AS102 arc in NGC 300 and the position of the X-ray
source P42 (X-ray isolines) overlayed (the NOT image by S.Larsen).
At the bottom right - the IC133 complex in M33 (DSS), with the position
of M33 X-4 source shown as the yellow circle.

Fig. 12. The left row: the Bubble complex in NGC 6946 (top, HST PC);
the round complex (~300 pc in size, around the bright cluster), 1.5' West
of the center of M51 (middle, HST PC); the round complex (the very center
of the image) inside  the Northern spiral arm of NGC 1232 (bottom, VLT).
The right row: the NE segment of the LMC, 30 Dor nebula is at the middle
bottom (top); the LMC4 region; the SE complex in M83 with two arcs
of clusters inside (VLT).

\end{document}